\documentclass[conference]{IEEEtran}
\usepackage{cite}
\usepackage{amsmath,amssymb,amsfonts}
\usepackage{algorithmic}
\usepackage{graphicx}
\usepackage{textcomp}
\usepackage{xcolor}
\usepackage{float}
\usepackage{subfigure}
\usepackage{epsfig}
\usepackage {setspace}
\def\BibTeX{{\rm B\kern-.05em{\sc i\kern-.025em b}\kern-.08em
    T\kern-.1667em\lower.7ex\hbox{E}\kern-.125emX}}
\begin{document}

\title{A Novel Non-Stationary Channel Emulator for 6G MIMO Wireless Channels}

\author{Yuan Zong\textsuperscript{1,2}, Lijian Xin\textsuperscript{2*}, Jie Huang\textsuperscript{1,2}, Cheng-Xiang Wang\textsuperscript{1,2*}\\
	\textsuperscript{1}National Mobile Communications Research Laboratory, School of Information Science and Engineering,\\
	Southeast University, Nanjing 210096, China.\\	
	\textsuperscript{2}Purple Mountain Laboratories, Nanjing 211111, China.\\
	\textsuperscript{*}Corresponding Authors\\
	
	Email: zong\_yuan@seu.edu.cn, xinlijian@pmlabs.com.cn, \{j\_huang, chxwang\}@seu.edu.cn}

\maketitle

\begin{abstract}
The performance evaluation of sixth generation (6G) communication systems is anticipated to be a controlled and repeatable process in the lab, which brings up the demand for wireless channel emulators. However, channel emulation for 6G space-time-frequency (STF) non-stationary channels is missing currently. In this paper, a non-stationary multiple-input multiple-output (MIMO) geometry-based stochastic model (GBSM) that accurately characterizes the channel STF properties is introduced firstly. Then, a subspace-based method is proposed for reconstructing the channel fading obtained from the GBSM and a channel emulator architecture with frequency domain processing is presented for 6G MIMO systems. Moreover, the spatial time-varying channel transfer functions (CTFs) of the channel simulation and the channel emulation are compared and analyzed. The Doppler power spectral density (PSD) and delay PSD are further derived and compared between the channel model simulation and subspace-based emulation. The results demonstrate that the proposed channel emulator is capable of reproducing the non-stationary channel characteristics.
\end{abstract}


\section{Introduction}
Numerous potential network architectures and technologies have been presented for 6G wireless communication systems \cite{ONtheRoad}. It is indispensable to evaluate the performance of these systems in link and network layers. Compared with the software simulation and field test, the wireless channel emulator can reproduce channel characteristics in a repeatable, controllable, and real-time manner in the lab, which will facilitate the test and validation cycles efficiently. 

For flat fading channel, three major methods have been used to emulate the wireless channel. The first method is known as the sum-of-sinusoids (SoS) method, which is widely used to emulate Rayleigh and Rice fading channels with isotropic scattering conditions \cite{SOS}. A special case of the SoS is the sum-of-cisoids (SoC) method. A SoC-based hardware emulator was proposed to generate channel fading for the non-isotropic scattering environment with asymmetrical Doppler PSD in \cite{SOC}. The second approach, named as filter-based method, is commonly implemented by filtering the uncorrelated Gaussian sequence. 
%The filter with finite impulse response (FIR) or infinite impulse response (IIR) is often called the Doppler shaping filter. In \cite{FIR}, the coefficients of the FIR shaping filter are obtained from a truncated and windowed version of the impulse response, which is the inverse fast Fourier transform (IFFT) of the square root of Doppler spectrum. 
In \cite{IIR}, the least $p$th-norm approximation was employed to approximate Jakes Doppler PSD with a infinite impulse response (IIR) filter. 
%The autoregressive (AR) channel emulator is employed to generate  channels which essentially uses an IIR filter with rational transfer function to shape the spectrum of Gaussian write noise \cite{b6}.  
%FIR-based Doppler spread emulator do not suffer from stability problems, but a high computational complexity may be encountered on account of many coefficients required  to closely approximate the desired spectrum. The IIR filter-based methods are generally of low complexity, however, they are subject to numerical instabilities. 
The third approach employs the inverse fast Fourier transform (IFFT) to replace filter-based method. The overlap-save (OLS) method was used in \cite{IFFT1}, \cite{IFFT2} to implement the IFFT-based fading channel emulator allowing for the real-time emulation of a continuous transmission. 

As the signal bandwidth becomes wider, the wireless channel will exhibit frequency-selective fading. The tapped-delay-line (TDL) based  channel emulation is the most common approach which is implemented by the finite impulse response (FIR) filter with time-varying coefficients \cite{TDL}. In \cite{TDLFIR1}, the channel impulse response (CIR) of the satellite navigation channel was reconstructed sparsely by multipath aggregation, which can be used for the realistic TDL-based emulation. In \cite{TDLFIR2}, the hardware emulator for the ray tracing (RT) simulated channel was proposed, which employed the TDL-based channel emulation framework with the delay alignment and tap reduction of the FIR filter. Frequency domain processing is an alternative method with unique advantages compared with the TDL-based channel emulation. Though the time domain filtering method is well adapted for the emulation of single-input single-output (SISO) channels, its complexity will quadratically raise as the antenna elements increase in MIMO systems, which results in the reduction of available filter taps per subchannel for higher-order MIMO emulation. The frequency domain emulation was implemented and compared with the time domain emulation in \cite{Fre1}, which showed that frequency domain processing exhibits an initial fixed computational consumption due to Fourier transforms, but it is more computationally efficient and scalable for larger MIMO arrays. 
%However, it has been shown that processing in frequency domain exhibits initial implementation complexity but its complexity increases at a reduced rate with the antenna array size increasing. 
%It is concluded that frequency-domain emulation is more computationally efficient than TDL-based emulator for larger MIMO arrays with long channel impulse response (CIR).

%GBSM is widely applied to model MIMO channel owing to its ability in characterizing the environment-specific STF correlation.
The GBSM is widely applied to characterize MIMO channels owing to its ability of modeling the fading channel with the transmitter (Tx) and receiver (Rx) antenna patterns separately.
 Moreover, the GBSM can serve as a basis for 6G standard channel models in the future due to its acceptable accuracy, generality, and moderate complexity \cite{b14}. In \cite{6GPCM}, a three dimensional (3D) GBSM was proposed to characterize the STF non-stationary properties of 6G wireless channels. However, limited geometry-based channel emulators have been reported especially for the STF non-stationary GBSM channel emulation. The subspace-based method was proposed in \cite{HuangTowards}, \cite{b16}. The small-dimensional space was spanned by sinusoids in \cite{HuangTowards} , which was essentially implemented by SoS. In \cite{b16}, the signal processing algorithm based on the software defined radio (SDR) was presented for the real-time channel emulator, which projected the time-limited and frequency-limited CTF on the discrete prolate spheroidal sequences. The channel emulator for time domain non-stationary channels was proposed in \cite{b17}, \cite{b18}, which was implemented on the field-programmable gate array (FPGA) by sum-of-frequency-modulation (SoFM) method.  The TDL-based architecture was employed in \cite{b16,b17,b18}, which may lead to high computational complexity for massive MIMO channels \cite{Fre1}. However, the channel emulators for GBSM using frequency domain processing are still missing. To fill the research gap, this paper aims to propose a subspace-based non-stationary MIMO channel emulator that employs frequency domain processing. The main novelties and contributions of this paper are \mbox{summarized as follows:}
\begin{enumerate}
	\item A subspace projection based method is presented for preprocessing non-stationary fading sequences to reconstruct the CTF on the FPGA.
	\item An efficient and scalable channel emulator architecture for 6G MIMO systems is proposed by using subspace expansion method and frequency domain processing.
	\item The CTF, Doppler PSD, and delay PSD of the channel model simulation are derived and compared with those of the subspace-based emulation to validate the performance of the reconstructed channels.
\end{enumerate}

The remainder of this paper is organized as follows. In Section \ref{Model}, the non-stationary GBSM for MIMO systems is described. In \mbox{Section \ref{Emul}}, the channel emulation scheme based on the subspace expansion and frequency domain processing for the non-stationary GBSM is presented. Section \ref{Vali} compares and discusses the simulation and emulation results. Finally, conclusions are drawn in Section \ref{Conc}.

\section{The 3D Non-Stationary GBSM}\label{Model}

\begin{figure}[tb]
	\centering\includegraphics[width=0.5\textwidth]{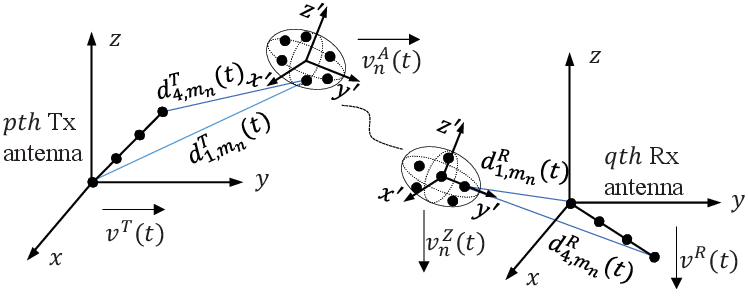}
	\caption{The 3D non-stationary GBSM for MIMO systems\cite{6GPCM}.}
	\label{modelfig}
\end{figure}

The GBSM is applied to model the small-scale fading channel between Tx antennas and Rx antennas \cite{6GPCM}, which is illustrated in Fig.~\ref{modelfig}. It is assumed that both the Tx and Rx are equipped with uniform linear arrays (ULAs) and omnidirectional antennas. The time-varying CIR between the $p$th Tx antenna and $q$th Rx antenna can be expressed by
\begin{align}\label{CIR}
%%	\begin{split}
		h_{qp}\left( {t,\tau} \right) =& \sum\limits_{n = 1}^{N_{qp}(t)} \sum\limits_{m = 1}^{M_n(t)}{\sqrt{P_{qp,m_n}(t)}} e^{j2\pi f_c\tau_{qp,m_n}(t)} \nonumber\\
		& \cdot\delta\left( {\tau - \tau_{qp,m_{n}}\left(t \right)} \right)
%%	\end{split}
\end{align}
where $\tau_{qp,m_n}(t)$ is the delay of the $m$th ray in the $n$th cluster between the $p$th Tx antenna and $q$th Rx antenna, $P_{qp,m_n}(t)$ is the power of the corresponding ray, $N_{qp}(t)$ is the number of cluster pairs and $M_n (t)$ is the number of scatterers in the $n$th cluster. To get the CIR of a MIMO channel with $P$ Tx antennas and $Q$ Rx antennas, it is obvious that up to $P\cdot Q\cdot M\cdot N$ rays need to be generated and the corresponding complex exponentials need to be calculated.

\subsection{Channel Parameter Calculation Methods}

The number of cluster pairs $N_{qp}(t)$  is a Markov process which represents the variation of clusters number. It can be obtained as the summation of survival clusters and newly generated clusters during $\Delta r$, $\Delta t$, and $\Delta f$, which is modeled by the birth-death processes to capture the STF non-stationarity of 6G wireless channels. 
%The survival clusters are determined by the survival probability $P_{surv}$. The number of newly generated clusters is generated according to a Poisson distribution randomly. 

The ray power $P_{qp,m_n}(t)$ depends on antenna, time, and frequency, which can be determined as \cite{6GPCM}
\begin{equation}\label{Power}
	P_{qp,m_n}(t) = \left( \frac{f_c + f}{f_c} \right)^{\gamma}\cdot  \exp\left( {- \frac{r_{\tau} - 1}{r_{\tau}\text{DS}} \tau_{qp,m_n}(t) } \right)
\end{equation}
where $r_{\tau}$ is the delay-dependent factor and $\gamma$ is a factor depended on frequency, which enables this model to capture the channel characteristic of large bandwidth scenarios.

The ray delay $\tau_{qp,m_n}(t)$ depends on the distance between Tx antenna and the scatterer in Tx-side cluster, Rx antenna and the scatterer in Rx-side cluster, and the virtual link between Tx-side cluster and Rx-side cluster.  $\tau_{qp,m_n}(t)$ can be given by
\begin{equation}\label{Delay}
	\tau_{qp,m_n}(t) = \frac{d_{p,m_n}^{T}(t) + d_{q,m_n}^{R}(t)}{c} + \tau_{link}.
\end{equation}

To obtain the time-varying distance such as $d_{p,m_n}^{T}(t)$ and $d_{q,m_n}^{R}(t)$, the initial coordinates and speeds of antennas and scatterers are required. The initial coordinates and speeds of antennas are deterministic, which can be configured according to the realistic scenario as well as the size and orientation of the antenna array. The scatterers coordinates and speeds are stochastic. Firstly, the ellipsoid Gaussian scattering distribution is used to generated the relative Cartesian coordinates $(x^{\prime},y^{\prime},z^{\prime})$ of scatterers relative to the center of the corresponding cluster. Then, the cluster center in the form of spherical coordinate $(d,\phi_E,\phi_A)$ is generated according to the proper random distribution.  In this model, $d$ is generated by exponential distribution and $\phi$ is generated by Gaussian distribution. Finally, the initial Cartesian coordinates $(x,y,z)$ of scatterers are obtained by the transformation in \cite{B5GCM}.
Since the parameters used for the distance calculation are generated, $d_{p,m_n}^{T}(t)$ can be determined as
\begin{align}\label{Distance}
		&d_{p,m_n}^{T}(t) = \nonumber\\ &\left\| \vec{C}_{m_n}^{A}(0) - \vec{A}_{p}^{T}(0) + \int_{0}^{t}{\left( {\vec{v}_{n}^{A}\left( t^{\prime} \right) - \vec{v}^{T}\left( t^{\prime} \right)} \right)dt^{\prime}} \right\|
\end{align}
where $\vec{C}_{m_n}^{A}(0)$ and $\vec{A}_{p}^{T}(0)$ are the initial Cartesian coordinates of the scatterer and antenna at Tx-side, $\vec{v}_{n}^{A}\left( t^{\prime} \right)$ and $ \vec{v}^{T}\left( t^{\prime} \right)$ are the time-varying speeds of cluster and Tx antenna. The calculation of $d_{q,m_n}^{R}(t)$ is similar.

\subsection{The STF Non-Stationary CTF}

The spatial time-varying non-stationary CTF $H_{qp}({t,f})$ is the Fourier transform of $h_{qp}({t,\tau})$ with respect to (w.r.t.) delay $\tau$. Therefore, $H_{qp}({t,f})$ can be determined as
\begin{equation}\label{CTF}
	H_{qp}\left( {t,f} \right) = {\sum\limits_{n = 1}^{N_{qp}(t)}{\sum\limits_{m = 1}^{M_{n}(t)}{\sqrt{P_{qp,m_{n}}(t)}e^{j 2\pi{({f_{c} - f})}\tau_{qp,m_{n}}{(t)} }}}}.
\end{equation}

The characteristics of massive MIMO channel can be captured by this model, since different antenna pairs correspond to different delays which results in different power. On one hand, different antenna elements have different angles of arrival or departure for the same scatterer, which shows the spherical wavefront characteristic of the channel. On the other hand, as the size of antenna array grows larger, different scatterers or clusters may be observed by different antenna pairs, which shows the space domain non-stationarity of \mbox{the massive MIMO channel.}

The Doppler shift of a ray can be obtained as the derivative of phase w.r.t. time, which can be expressed as
\begin{equation}\label{Doppler0}
	f_{qp,m_n}(t) = \frac{d\tau_{qp,m_n}(t)}{dt}f_c = \frac{d\left(d_{p,m_n}^{T}(t)+d_{q,m_n}^{R}(t)\right)}{dt}\frac{1}{\lambda}
\end{equation}
where $\lambda$ is the wavelength. When the Tx remains fixed and the speed of Rx is constant, the derivative of distance is approximated as
\begin{align}\label{Doppler2}
%%	\begin{split}
	{d\left( {d_{{q,m}_{n}}^{R}(t)} \right)}/{dt} &\approx - {\cos\left( \omega_{q,m_{n}}^{R} \right)}v^{R} \nonumber\\
	&+ \frac{{\sin}^{2}\left( \omega_{q,m_{n}}^{R} \right)(v^{R})^2t}{\left\lbrack {D_{m_{n}}^{R} - {\cos\left( \theta_{m_{n}}^{R} \right)}\left( {q - 1} \right)\delta^{R}} \right\rbrack}
%%	\end{split}
\end{align}
where $\cos\left( \theta_{m_{n}}^{R} \right)$ and $\cos\left( \omega_{q,m_{n}}^{R} \right)$ can be calculated according to \cite{B5GCM}, $\delta^{R}$ is the antenna spacing at the Rx.
In this case, the Doppler frequency can be approximated as
\begin{align}\label{Doppler3}
%%	\begin{split}
	f_{qp,m_{n}}(t) &\approx \frac{- {\cos\left( \omega_{q,m_{n}}^{R} \right)}v^{R}}{\lambda}  \nonumber\\
		&+ \frac{{\sin}^{2}\left( \omega_{q,m_{n}}^{R} \right)(v^{R})^2t}{\lambda\left\lbrack {D_{m_{n}}^{R} - {\cos\left( \theta_{m_{n}}^{R} \right)}\left( {q - 1} \right)\delta^{R}} \right\rbrack}.
%%	\end{split}
\end{align}
It can be seen that the instantaneous Doppler frequency varies with time which means that the channel model is able to capture the time domain non-stationary characteristic for high-mobility scenarios. Moreover, it also indicates that the Doppler frequency can be approximated by a linear variation function, which provides a theoretical inspiration for the following subspace expansion method.

\section{Channel Emulation Based on Subspace Expansion}\label{Emul}

To emulate the effect of wideband wireless channels, the input signal of a channel emulator needs to be spread in frequency domain and time domain. When considering MIMO wireless communication systems, the spatial correlation between subchannel fading also needs to be emulated. By reconstructing the CTF generated according to the 3D non-stationary GBSM on the hardware platform such as the FPGA, the STF correlation properties of various modeled wireless channels can be emulated.

When the lowpass equivalent $s_p (t)$ of transmitted signal is emitted from the $p$th Tx antenna, the complex envelope $r_q (t)$ of received signal at the $q$th Rx antenna can be given by
\begin{equation}\label{IOequ}
	r_{q}(t) = \sum\limits_{p = 1}^{P}{\int_{- \infty}^{\infty}{S_{p}(f)H_{qp}\left( {t,f} \right)e^{j2\pi ft}df}}
\end{equation}
where $S_p (f)$ is the Fourier transform of $s_p (t)$. The equivalent signal processing of \eqref{IOequ} needs to be implemented by the channel emulation.

\subsection{Channel Emulation Architecture For GBSM}\label{AA}
The SDR-based channel emulator architecture is presented for the non-stationary GBSM. For the GBSM, the parameter configuration is very elaborated and Frobenius norm calculation is necessary to obtain the delay that is essential for the CTF generation. Therefore, it is sensible to implement the parameter configuration and CTF generation of the GBSM on a personal computer (PC). Then, the preprocessing of the original CTF is performed on the PC in order to reconstruct the CTF with low hardware complexity in real time on the FPGA. Moreover, it will reduce the data transmission bandwidth between the FPGA and the PC owing to the enormous volume of CTF matrix especially for the massive MIMO and high-mobility scenarios.
 
\begin{figure}[tb]
	\centering\includegraphics[width=0.44\textwidth]{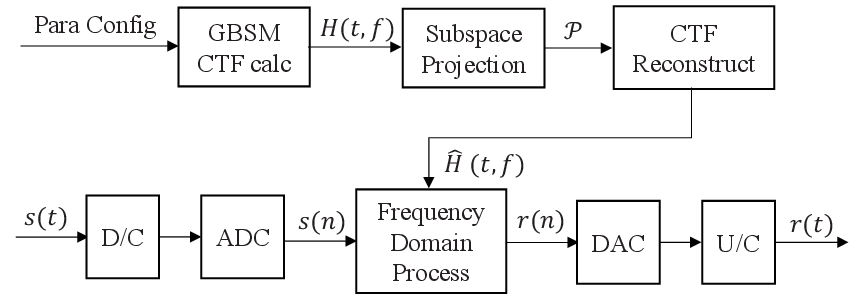}
	\caption{Channel emulator structure for GBSM.}
	\label{emufig}
\end{figure}

The SDR platform consists of radio frequency frontend and baseband processing unit. The down-converter(D/C) is used to convert the passband input signal to the baseband. The digital baseband signal $s(n)$ is than obtained by the analog-to-digital converter (ADC). On the FPGA, the result of preprocessing denoted as $\mathcal{P}=(\mathbf{A},\alpha_k,\beta_k)$ is transmitted to it and stored in the memory. The CTF with continuous time evolution is reconstructed from $\mathcal{P}$. FFT/IFFT and multiply-accumulate (MAC) operation with the reconstructed CTF are performed to process input signals. Then, the digital baseband signal $r(n)$ corrupted by the CTF is converted to bandpass signal through the digital-to-analog converter (DAC) and up-converter(U/C). Fig.~\ref{emufig} illustrates the channel emulator structure for the non-stationary GBSM.

\begin{figure}[tb]
	\centering\includegraphics[width=0.32\textwidth,height=0.25\textwidth]{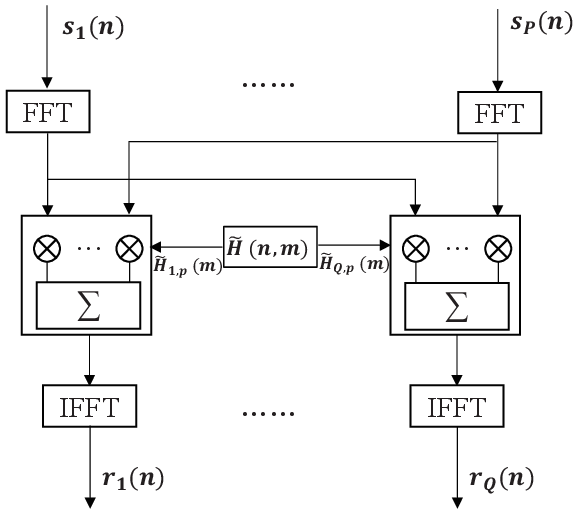}
	\caption{Frequency domain processing diagram.}
	\label{Freprofig}
\end{figure}

\subsection{Project of Original CTF to Subspace}\label{BB}
By substituting \eqref{Doppler2} into \eqref{Delay}, the subchannel CTF will be approximated as
\begin{equation}
	H\left( {t,f} \right) \approx {\sum\limits_{l = 1}^{L(t)}{\eta_{l}e^{- \xi_{l}(t_{l} + a_{l}t + b_{l}t^{2})}e^{j2\pi(f_{c} - f)(t_{l} + a_{l}t + b_{l}t^{2})}}}
\end{equation}
where $L(t) = \sum\nolimits_{n = 1}^{N(t)}M_n(t)$, $t_{l}$, $\eta_{l}$, $\xi_{l}$, $a_{l}$, and $b_{l}$ are constants independent of time and frequency. For example, the typical values of $M_n(t)$ and $N(t)$ in the clustered delay line (CDL) models of \cite{b20} are fixed as 20 and 23. $L(t)$ is a very large value using the above parameters. It makes the channel model computationally heavy to implement on the FPGA, especially for MIMO channel emulation. To solve this problem, $H(t,f)$ can be projected on a $K$ dimensional subspace since the CTF matrix is sparse in Doppler domain and delay domain.

The channel fading w.r.t time $t$ for different frequency bins $f_i$ can be defined as
\begin{equation}
	\mathbf{h} = \left\lbrack H\left( {t,f_{1}} \right),H\left( {t,f_{2}} \right),\cdots,H\left( {t,f_{I}} \right) \right\rbrack
\end{equation}
where $H(t, f_i )$, $i =1, … ,I$, are complex Gaussian process in general. Chirp signals satisfy the restricted isometry property (RIP) to reconstruct Gaussian stochastic process perfectly \cite{b21}. Besides, they can be implemented fastly on the FPGA, which leads to rapid channel fading reconstruction. Therefore, the $K$ dimensional subspace can be spanned by $K$ chirp signals. The $k$th basis of the $K$ dimensional subspace can be expressed as
\begin{equation}
	b_k(t)=e^{j2\pi(\alpha_{k}t + \beta_{k}\frac{t^{2}}{2})},k =1, … ,K.
\end{equation}
Here, $\alpha_k$ is the initial frequency and $\beta_k$  is the chirp rate, which can be obtained by the statistical distribution of parameters in \eqref{Doppler3} and the method in \cite{b21}. Based on the assumption in \eqref{Doppler3}, the range of $\beta_k$ and $\alpha_k$ can \mbox{be approximated as}
\begin{align}
%%	\begin{split}
		\beta_{k} \in& \left[ \min\left\{ \frac{{\sin}^{2}\left( \omega_{q,m_{n}}^{R} \right)(v^{R})^2}{\lambda\left\lbrack {D_{m_{n}}^{R} - {\cos\left( \theta_{m_{n}}^{R} \right)}\left( {q - 1} \right)\delta^{R}} \right\rbrack} \right\},  \right. \nonumber\\
		& \left. \max\left\{\frac{{\sin}^{2}\left( \omega_{q,m_{n}}^{R} \right)(v^{R})^2}{\lambda\left\lbrack {D_{m_{n}}^{R} - {\cos\left( \theta_{m_{n}}^{R} \right)}\left( {q - 1} \right)\delta^{R}} \right\rbrack}\right\}  \right]
%%	\end{split}
\end{align}
\begin{equation}
	\alpha_{k} \in \left[ \min\left\{ \frac{- {\cos( \omega_{q,m_n}^{R})}v^{R}}{\lambda} \right\}, \max\left\{ \frac{- {\cos( \omega_{q,m_n}^{R})}v^{R}}{\lambda} \right\}\right].
\end{equation}

To obtain channel fading by the linear combination of this set of basis vectors, the projection of $\mathbf{h}$ to each chirp signal needs to be pre-calculated. The matrix composed of chirp signals is defined as $\mathbf{\Psi}=  \left\lbrack b_1(t),b_2(t),\cdots,b_k(t)  \right\rbrack$. We assume that $\mathbf{\Psi}$ can be orthogonalized as $\mathbf{\Phi}$ by matrix $\mathbf{G}$, which can be represented as
\begin{equation}
	\mathbf{\Phi} = \mathbf{\Psi} \cdot \mathbf{G}
\end{equation}
where $\mathbf{G}$ is an upper triangular matrix and can be given by
\begin{equation}
	\mathbf{G} = \begin{bmatrix}
		1 & \cdots & c_{1K} \\
		\vdots & \ddots & \vdots \\
		0 & \cdots & 1
	\end{bmatrix}^{- 1}.
\end{equation}
Here,
\begin{equation}
	c_{jk} = \frac{<b_{k}(t),g_{j}(t)>}{<g_{j}(t),g_{j}(t)>},\quad j<k
\end{equation}
where $<\cdot ~,~\cdot>$ denotes the inner product operation, $b_k(t)$ and $g_j(t)$ represent the $k$th and $j$th column vectors of $\mathbf{\Psi}$ and $\mathbf{\Phi}$, respectively. Since $\mathbf{\Phi}$ are orthogonal, the projection $\mathbf{X}$ of $\mathbf{h}$ on $\mathbf{\Phi}$ can be calculated as
\begin{equation}	
	x_{ji} = \frac{<H(t, f_i ),g_j(t)>}{< g_j(t),g_j(t) >}
\end{equation}
where $x_{ji}$ is an element in $j$th row and $i$th column of $\mathbf{X}$. After $\mathbf{G}$ and $\mathbf{X}$ are obtained for $\mathbf{\Phi}$ and $\mathbf{\Psi}$, the projection $\mathbf{A}$ of $\mathbf{h}$ on $\mathbf{\Psi}$ can be calculated as $\mathbf{A} = \mathbf{G} \cdot \mathbf{X}$.
%\begin{equation}	
%	A = G \cdot X.
%\end{equation}
Then, the channel fading can be reconstructed as
\begin{equation}	
	\mathbf{\hat{h}} = \mathbf{\Psi} \cdot \mathbf{A}.
\end{equation}

Hence, $\mathbf{\Psi}$ can be used as $K$ bases to reconstruct channel fading $\mathbf{h}$ in time domain during a time window $T_w$. The CTF approximated by the subspace expansion approach can be obtained as
\begin{equation}\label{reCTF}
	\hat{H}\left( {t,f_i} \right) = \sum\limits_{k = 1}^{K}{a_{qp,k}(f_i) \cdot e^{j2\pi(\alpha_{k}t + \beta_{k}\frac{t^{2}}{2})}}, t \in [t_0, t_0+T_w]
\end{equation}
where $a_{qp,k}(f_i)$ is an element in $k$th row and $i$th column of $\mathbf{A}$.
Since $(\mathbf{A},\alpha_k,\beta_k)$ is sent and stored on the FPGA,  $e^{j2\pi(\alpha_{k}t + \beta_{k}\frac{t^{2}}{2})}$ and $\hat{H}({t,f})$ can be generated efficiently by using time division SoFM method on the FPGA \cite{b17}.

\begin{figure*}[htbp]
	\subfigure[]{
		\centering\includegraphics[width=0.65\textwidth]{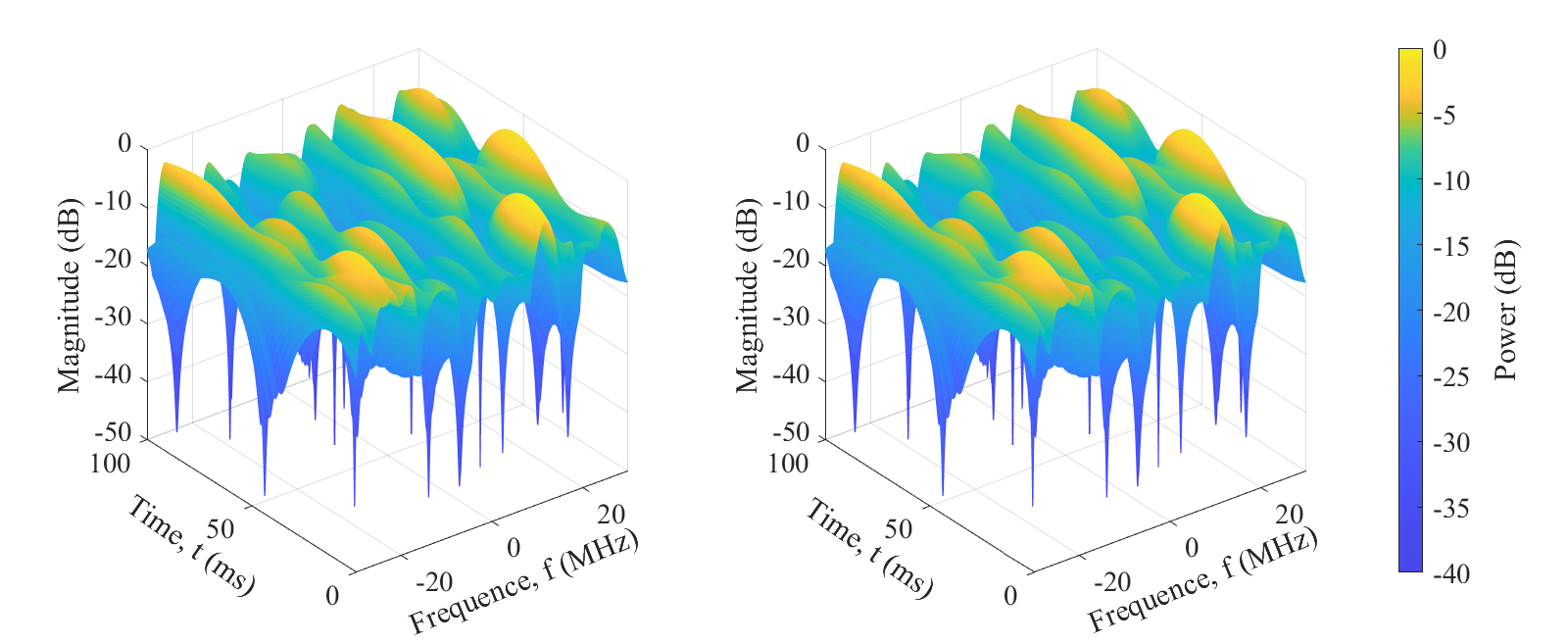}}
	\subfigure[]{
		\centering\includegraphics[width=0.3\textwidth, height=0.25\textwidth]{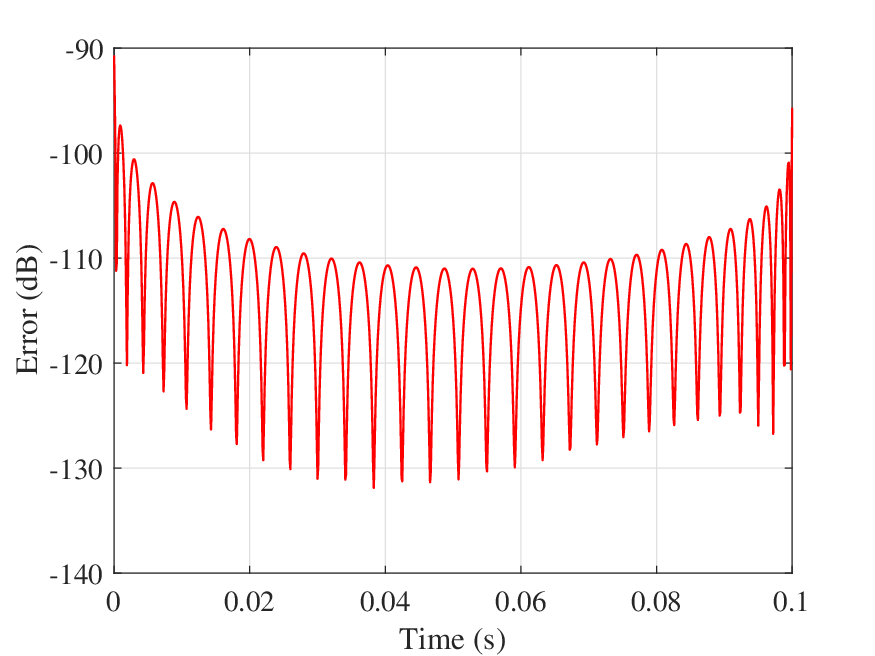}}
	\caption{(a) Original and reconstructed CTF, (b) CTF error ($f_c$=2.6 GHz, $B$=60 MHz, $N$=23, $M$=20, $K$=30).}
	\label{CTFfig}
\end{figure*}

\subsection{Frequency Domain Processing}
When $\hat{H}_{qp}(t,f)$ is reconstructed on the FPGA platform, the channel frequency response (CFR) $H_{qp}(f)$ of the channel can be obtained directly from $\hat{H}_{qp}(t,f)$ during channel sample period $T_{ch}$. It means $H_{qp}(f)$ will automatically update every $T_{ch}$, which can characterize the realistic channels better. The discrete CFR $H_{qp}(m)$ with $N_H$ points is predefined in the channel model by sampling $H_{qp}(f)$. The CIR $h_{qp}(\tau)$, which exhibits the maximum delay spread $\tau_m$, is the inverse Fourier transform of $H_{qp}(f)$. To compute linear convolution using circular convolution which is equivalent to frequency domain multiplication \cite{Fre1}, a vector $s_{p,i}(n)$ of $N_s=N_H-N_a$ points is taken from input signal $s_p(n)$, where $N_a=\lfloor\tau_m⁄T_s\rfloor$, $\lfloor \cdot \rfloor$ denotes rounding down, and $T_s$ is the signal sample period, $T_s=T_{ch} / N_s$. Then, an FFT of length $N_H$ is performed with $s_{p,i}(n)$  by appending $N_a$ zeros at the end of $s_{p,i}(n)$  to obtain $S_{p,i} (m)$.  For the $P \times Q$ MIMO system, the received frequency domain signal can be determined  as 
\begin{equation}\label{FreIO}	
	R_{q,i}(m) = \sum\limits_{p = 1}^{P}{S_{p,i}(m)H_{qp}}(m),1 \leq q \leq Q.
\end{equation}

Then, an IFFT of length $N_H$ is performed with $R_{q,i}(m)$ to obtain $r_{q,i}(n)$. The frequency domain processing diagram is presented in Fig.~\ref{Freprofig}. The overlap-and-add method is applied to overcome the discontinuity issue inherently in the FFT/IFFT so that continuous streaming output signal of the emulator can be obtained as $r_q (n) = \sum\limits_{i}{r_{q,i}\left( {n - iN_{s}} \right)}$.
%\begin{equation}	
%	r_{q}(n) = \sum\limits_{i}{r_{q,i}\left( {n - iN_{s}} \right)}.
%\end{equation}

\section{Simulation and Validation}\label{Vali}

\subsection{CTF}
The CTF of the GBSM in \eqref{CTF} is simulated and compared with the CTF reconstructed according to the method of section \ref{BB}. In the simulation, the carrier frequency $f_c$ is 2.6 GHz, and the initial positions of Tx and Rx are set as $A_1^T (0)=(\text{0},\text{0},\text{35})$~m and $A_1^R (0)=(\text{10},\text{0},\text{1.5})$~m. The speed of Rx $v^R$ is $(\text{10}, \text{0}, \text{0})$~m/s and the Tx remains fixed. The other related parameters of the GBSM are set according to \cite{6GPCM}. The number of bases in the subspace-based emulation is set as $K=30$.

The mean error between the normalized CTFs generated from the GBSM and subspace reconstruction is defined as
\begin{equation}	
	e(t) = \frac{1}{B}{\int_{- B/2}^{B/2}{\left( {H\left( {t,f} \right) - \hat{H}\left( {t,f} \right)} \right)df}}.
\end{equation}
Here, $B$ is the channel bandwidth, which is set as 60 MHz in the simulation. The magnitude values of the original CTF $H({t,f})$ and reconstructed CTF  $\hat{H}({t,f})$, which characterize the time-selective and frequency-selective properties of the subchannel between the $4$th Tx antenna and 1st Rx antenna,  are shown in Fig.~\ref{CTFfig}(a). The error function $e(t)$ is shown in Fig.~\ref{CTFfig}(b). The subchannel CTF obtained from subspace reconstruction provides a quite good approximation to the original one generated from the GBSM with an error smaller than -90 dB. Therefore, the MIMO channel can be emulated with all reconstructed subchannels of different antenna pairs.

\begin{figure}[tb]
	\centering
	\includegraphics[width=0.44\textwidth]{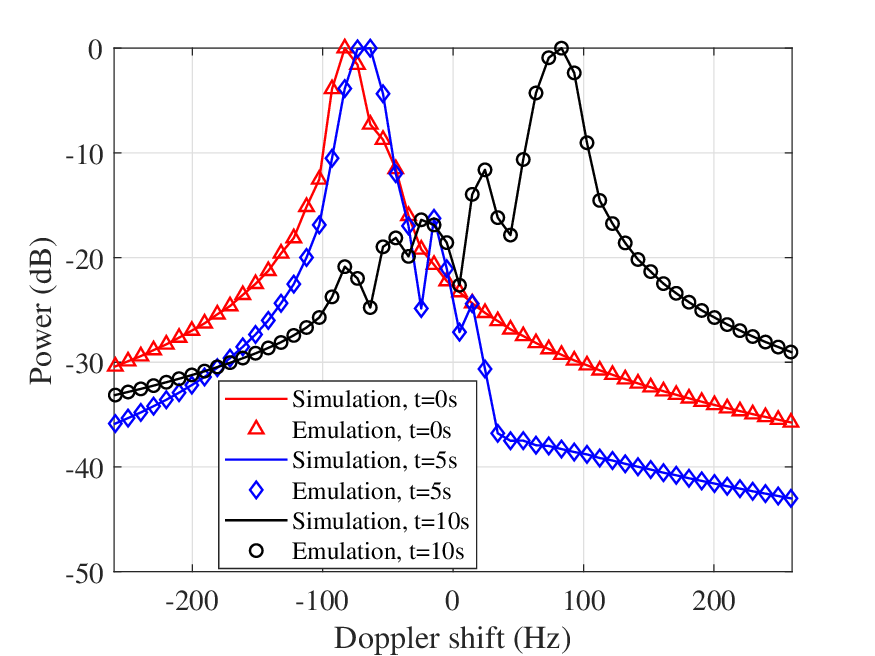}
	\caption{Comparisons of Doppler PSDs at different time instants ($f_c$=$\text{2.6}$~GHz, $A_1^T (0)$=$(\text{0},\text{0},\text{35})$~m, $A_1^R (0)$=$(\text{10},\text{0},\text{1.5})$~m, $v^R$=$(\text{10}, \text{0}, \text{0})$~m/s).}
	\label{DPSDfig}
\end{figure}

\subsection{Doppler PSD}
The Doppler PSD is defined as the short-time Fourier transform of the temporal autocorrelation function (ACF) for the time domain non-stationary channel model, and it can be obtained as
\begin{equation}	
	S_{qp}\left( {t,v} \right) = {\int{R_{qp}(t,{\Delta}t)w(t - \Delta t)e^{- j2\pi v\Delta t}d\Delta t}}
\end{equation}
where $w(t - \Delta t)$ is the window function and $R_{qp}(t,{\Delta}t)$ is the temporal  ACF at zero-frequency point.

The output Doppler PSD of the channel emulator is compared with simulated ones at $t=\text{0}$~s, $t=\text{5}$~s, and $t=\text{10}$~s, which are shown in Fig.~\ref{DPSDfig}. The variation of Doppler PSD indicates the inconsistency of temporal ACF at these time regions, which reflects the time domain non-stationarity of the channel model simulation and emulation. Meanwhile, the emulation results are well matched with the theoretical simulation, which shows that the channel emulator is able to reproduce the channel characteristic in time domain.

\subsection{Delay PSD}
The delay PSD can be obtained by the inverse Fourier transform of $R_{qp}(t,{\Delta}f)$ and can be determined as
\begin{equation}	
	S_{qp}\left( {t,\tau} \right) = {\int{R_{qp}(t,{\Delta}f)e^{j2\pi\tau\Delta f}d\Delta f}}
\end{equation}
where $R_{qp}(t,{\Delta}f)$ is the frequency correlation function.

\begin{figure}[tb]
	\subfigure[]{
		\centering\includegraphics[width=0.22\textwidth, height=0.2\textwidth]{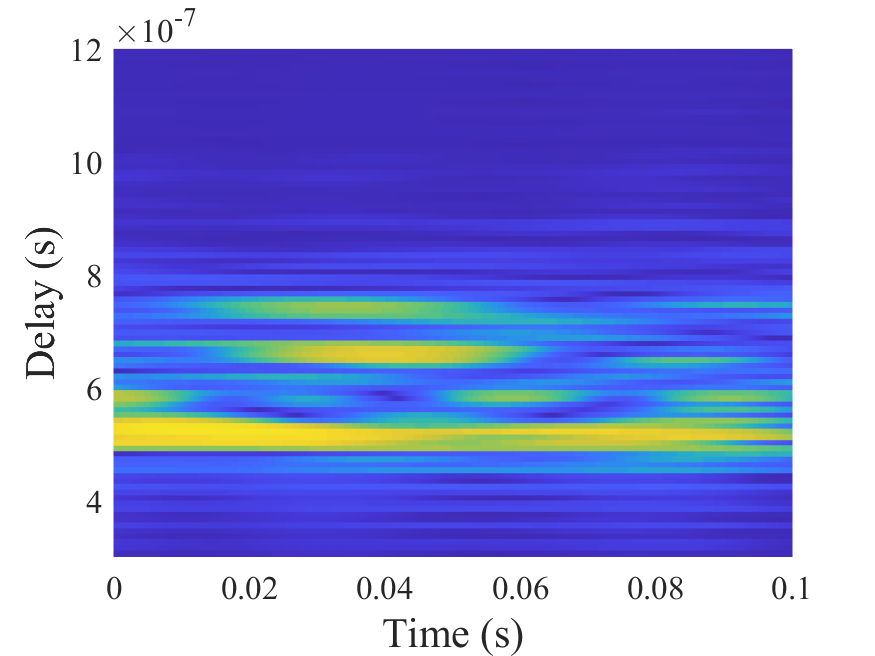}}\subfigure[]{
		\centering\includegraphics[width=0.24\textwidth, height=0.2\textwidth]{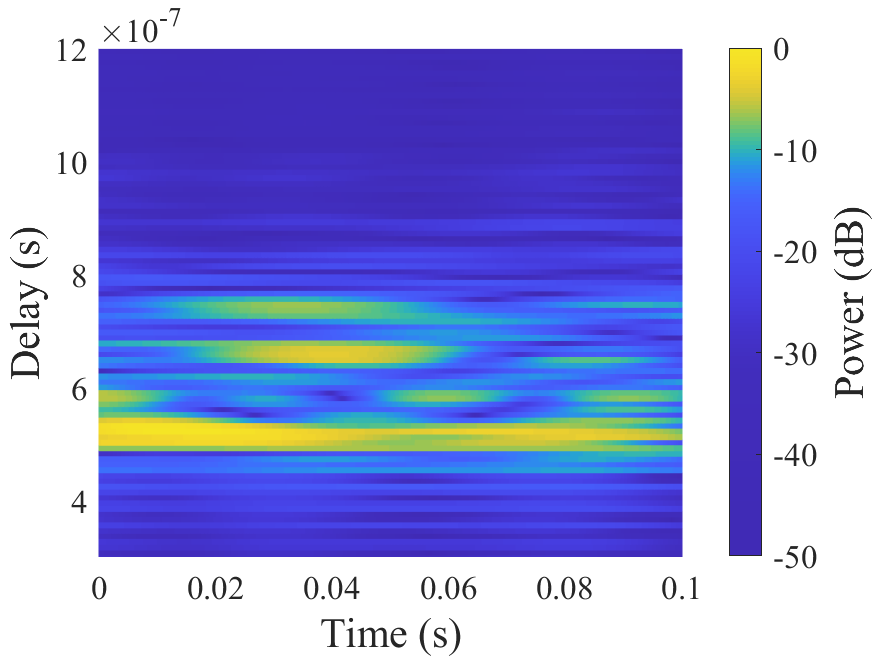}}
	\caption{(a) Simulated delay PSD and (b) Emulated delay PSD ($f_c$=$2.6$~GHz, $A_1^T (0)$=$(\text{0},\text{0},\text{35})$~m, $A_1^R (0)$=$(\text{100},\text{0},\text{1.5})$~m, $ v^R$=$(\text{10}, \text{0}, \text{0}) $~m/s).}
	\label{PDPfig}
\end{figure}

The simulated and emulated delay PSDs are given in Fig.~\ref{PDPfig}. The delay spread indicates the frequency-selectivity of the multipath channel and the time-varying characteristic of delay PSD exhibits the birth and death processes of valid rays over time. It is shown that the emulated and simulated delay PSDs are well matched, which indicates that the channel emulator is capable of reproducing the aforementioned channel characteristics.

\section{Conclusions}\label{Conc}
In this paper, a twin-cluster GBSM that can capture STF non-stationary channel characteristics has been introduced  for 6G MIMO communication systems. Based on the channel model, a subspace expansion approach has been presented to reconstruct the non-stationary CTF and a frequency domain processing scheme has been employed to implement MIMO channel emulation. The channel emulator architecture has been presented with the subspace projection preprocessing on the PC and the frequency domain processing using the reconstructed CTF on the FPGA. The simulated time-varying CTF, Doppler PSD, and delay PSD have been compared with the emulated ones, which shows well match between them. Therefore, the proposed channel emulator can replicate non-stationary channel characteristics with a negligible error to facilitate the test of 6G MIMO systems.

\section*{Acknowledgment}
\begin{spacing}{0.95}
This work was supported by the National Natural Science Foundation of China (NSFC) under Grants 61960206006, 62394290, 62394291, and 62271147, the Fundamental Research Funds for the Central Universities under Grant 2242022k60006, the Key Technologies R\&D Program of Jiangsu (Prospective and Key Technologies for Industry) under Grants BE2022067, BE2022067-1, BE2022067-3, and BE2022067-4, the EU H2020 RISE TESTBED2 project under Grant 872172, the High Level Innovation and Entrepreneurial Doctor Introduction Program in Jiangsu under Grant JSSCBS20210082, the Start-up Research Fund of Southeast University under Grant RF1028623029, and the Fundamental Research Funds for the Central Universities under Grant 2242023K5003.
\end{spacing}

\end{document}